\documentclass[5p]{elsarticle}
\usepackage{graphicx}
\usepackage{color}
\usepackage{bm}
\usepackage{multirow}
\usepackage{epsfig}
\usepackage[english]{babel}

\newcommand{\bc}           {\begin{center}}
\newcommand{\ec}           {\end{center}}
\newcommand{\bq}           {\begin{eqnarray}}
\newcommand{\eq}           {\end{eqnarray}}
\newcommand{\be}           {\begin{equation}}
\newcommand{\ee}           {\end{equation}}
\newcommand{\bi}           {\begin{itemize}}
\newcommand{\ei}           {\end{itemize}}
\newcommand{\er}           {$\pm$}

\begin{document}
\begin{frontmatter}

\title{\boldmath\bf $N^{\bf *}$ decays to $N\omega$ from new data on $\gamma p\to \omega p$
\vspace{5mm}
}

\author[HISKP]{I.~Denisenko}
\author[HISKP,GATCHINA]{A.V.~Anisovich}
\author[FSI]{V. Crede}
\author[PI]{H. Eberhardt}
\author[HISKP]{E.~Klempt}
\author[HISKP,GATCHINA]{V.A.~Nikonov}
\author[HISKP,GATCHINA]{A.V.~Sarantsev}
\author[PI]{H.~Schmieden}
\author[HISKP]{U.~Thoma}
\author[HISKP]{A.~Wilson}

\address[HISKP]{Helmholtz-Institut f\"ur Strahlen- und Kernphysik der
Universit\"at Bonn, Germany}
\address[GATCHINA]{Petersburg
Nuclear Physics Institute, Gatchina, Russia}
\address[FSI]{Department of Physics, Florida State University, Tallahassee, USA}
\address[PI]{Physikalisches Institut  der Universit\"at Bonn, Germany}



\begin{abstract}
Data on the reaction $\gamma p\to \omega p$  with $\omega\to\pi^0\gamma$, taken
with unpolarized or polarized beams in combination with an unpolarized or
polarized proton-target, were analyzed within the Bonn-Gatchina (BnGa) partial wave analysis.
Differential cross sections, several spin
density matrix elements, the beam asymmetry $\Sigma$, the normalized
helicity difference $E$, and the correlation $G$ between linear
photon and longitudinal target polarization were included in a large data
base on pion and photo-induced reactions. The data on $\omega$ photoproduction
are used to determine twelve $N^*\to N\omega$ branching ratios; most
of these are determined for the first time.
\vspace{2mm}   \\
{\it PACS: 11.80.Et,  13.30.-a,  13.40.-f, 13.60.Le\vspace{2mm}  \\ }
Submitted: \today
\end{abstract}

\vskip 5mm

\end{frontmatter}

\section{Introduction}
The interaction of real or virtual photons with protons at high energies - as studied
extensively at HERA for masses up to 250\,GeV \cite{Wolf:2009jm} - is successfully
described as a diffractive process. The photon converts into a
vector meson ($\rho^0$, $\phi$, $\omega$) of the same quantum
numbers $J^{PC} = 1^{--}$, i.e. of
identical spin, parity, and charge parity.
The vector meson then scatters off the proton
by the exchange of Pomerons, virtual color- and
flavorless objects carrying the quantum numbers of the vacuum
\cite{Donnachie:1994zb,Donnachie:1999yb}.  A detailed comparison of
the photoproduction of $\rho^0$, $\phi$, and $\omega$ mesons reveals,
however, that for $\omega$ photoproduction at intermediate energies,
$E_\gamma < 5$\,GeV, Pomeron exchange is no longer sufficient to
reproduce the data, and it has been suggested that pion and $f_2$ exchange
become the dominant contributions \cite{Sibirtsev:2003qh}.

At lower energies, close to the $\omega$  production
threshold, $N^*$ resonances are likely to contribute to the
reaction. The SAPHIR collaboration reported differential
cross sections and spin density matrix elements in the center-of-mass energy range
from the $\omega$ production threshold to 2.4\,GeV \cite{Barth:2003kv}.
The authors concluded that in this mass range
diffraction is no longer dominant, and that resonance formation must
play an important role. GRAAL data on this reaction confirmed the
need for resonances to understand the dynamics of $\omega$
photoproduction \cite{Ajaka:2006bn}. The CBELSA/TAPS collaboration reported large
photon asymmtries which indicated s-channel resonance formation on top of
t-channel exchange processes~\cite{Klein:2008aa}. The CLAS
collaboration reported a high-statistics study of $\omega$
photoproduction and analyzed the data with a partial-wave-analysis
model~\cite{Williams:2009ab,Williams:2009aa}. Differential cross
sections and spin density matrix elements were described with
reasonable accuracy when several resonances were introduced:
$N(1680)5/2^+$ and $N(1700)3/2^-$ near threshold and at least one
higher-mass state, $N(2190)$ $7/2^-$. Suggestive evidence was reported
for the presence of a $J^P = 5/2^+$ state around 2\,GeV. The $J^P =
3/2^+$ wave was reported to have a complicated structure, possibly with two
close-by resonances in the 1.9\,GeV region. Recently, photoproduction
of $\omega$ mesons off the proton was studied by the A2 Collaboration at
MAMI, and differential cross sections were presented from threshold to
$E_\gamma  = 1.4$\,GeV with 15-MeV binning and full angular coverage
\cite{Strakovsky:2014wja}. No resonant contributions were discussed.

Partial wave analyses confirmed the need for nucleon excitations to describe
photoproduction of $\omega$ mesons. Qiang Zhao \cite{Zhao:2000tb} used an effective
Lagrangian and found that $N(1720)3/2^+$ and $N(1680)5/2^+$ dominate the reaction.
Predictions of Capstick and Roberts \cite{Capstick:2000qj}
were used in \cite{Oh:2000zi} to calculate the $\omega$ photoproduction cross section.
The resonant contributions were shown to have a significant impact on the predictions.
Titov and Lee \cite{Titov:2002iv} applied an effective Lagrangian approach to study
the role of the nucleon resonances in $\omega$
photoproduction at energies near the threshold and found that their contribution is very
significant. In a pioneering coupled-channel analysis, Penner and Mosel \cite{Penner:2002ma} fitted data
on pion and photo-induced reactions including $\pi^-p\to\omega n$ 
\cite{Danburg:1971ui,Binnie:1974cq,Keyne:1976tj,Karami:1979ib}
and $\gamma p\to \omega p$ \cite{Barth:2003kv} and determined first
$N^*\to N\omega$ branching ratios. In a coupled-channel analysis including further data,
Shklyar {\it et al.} \cite{Shklyar:2004ba} found strong contributions from $N(1680)5/2^+$ 
and $N(1675)5/2^-$ to the $\gamma p\to \omega p$ reactions.

In this letter we report on a partial wave analysis of $\omega$
photoproduction of data taken recently at the Bonn {\it ELectron Stretcher
Accelerator} ELSA. We restrict the analysis on data from the CBELSA/TAPS
experiment; a discussion of discrepancies between different data sets and
the data dependence of the results will be presented elsewhere \cite{Anisovich:2016tbd}.
CBELSA/TAPS data on the differential cross section and on the
Spin-density Matrix Elements (SDMEs)  were reported
in \cite{Wilson:2015uoa}:
$\rho_{00}$, $\rho_{10}$, $\rho_{1\,-1}$
for unpolarized incident photons, and $\rho^1_{00}$, $\rho^1_{11}$, $\rho^1_{1\,-1}$,
$\rho^1_{10}$, $\rho^2_{10}$, $\rho^2_{1\,-1}$ for linearly
polarized photons. Differential cross sections and SDMEs cover the
photon energy range from 1150 to 2500\,MeV; the SDMEs for polarized
incident photons are restricted
to $E_\gamma < 1650$\,MeV. The SDMEs describe the polarization state
and the polarization transfer of the $\gamma p$ system to the final
state. Results on the beam asymmetry $\Sigma$ with respect to the
$\omega$ direction and with respect to the direction of the $\pi^0$
from $\omega\to\pi^0\gamma$ ($\Sigma_\pi$) are taken from \cite{Klein:2008aa}.
In \cite{Eberhardt:2015lwa}, the helicity asymmetry $E
=(\sigma_{1/2}-\sigma_{3/2})/(\sigma_{1/2}+\sigma_{3/2})$ was presented
for the photon energy range from 1108 to 2300\,MeV; the correlation
between linear photon polarization and transverse target polarization
($G$ and $G_\pi$) was given for one bin in photon energy covering 1108 to
1300\,MeV.

\section{Data from CBELSA/TAPS on $\gamma p\to \omega p$}

\begin{figure}[pt]
\centering
 \includegraphics[width=0.48\textwidth,height=0.5\textheight]{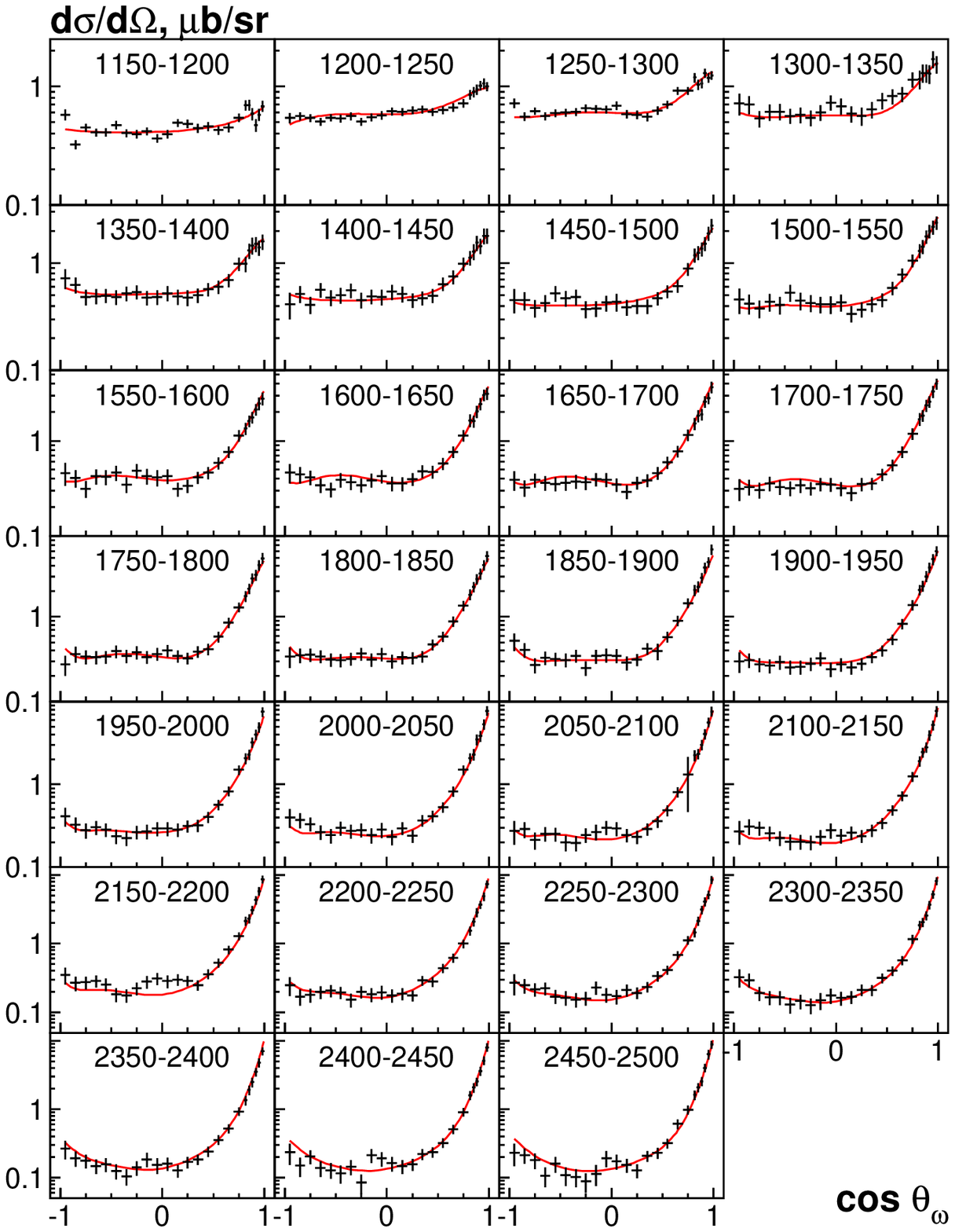}
\caption[Excitation functions]{\label{fig:Dcross_EF}(Color Online)
Differential Cross Sections for $\gamma p \rightarrow p \omega$ from
the CBELSA/TAPS experiment in bins of the photon
energy (in MeV) \cite{Wilson:2015uoa}. The total uncertainty
for each data point -- calculated from
the squared sum of statistical and systematic errors --
is represented as a vertical bar. The solid line  is the PWA
fit to the data. \vspace{3mm}}
\centering
\includegraphics[width=0.48\textwidth]{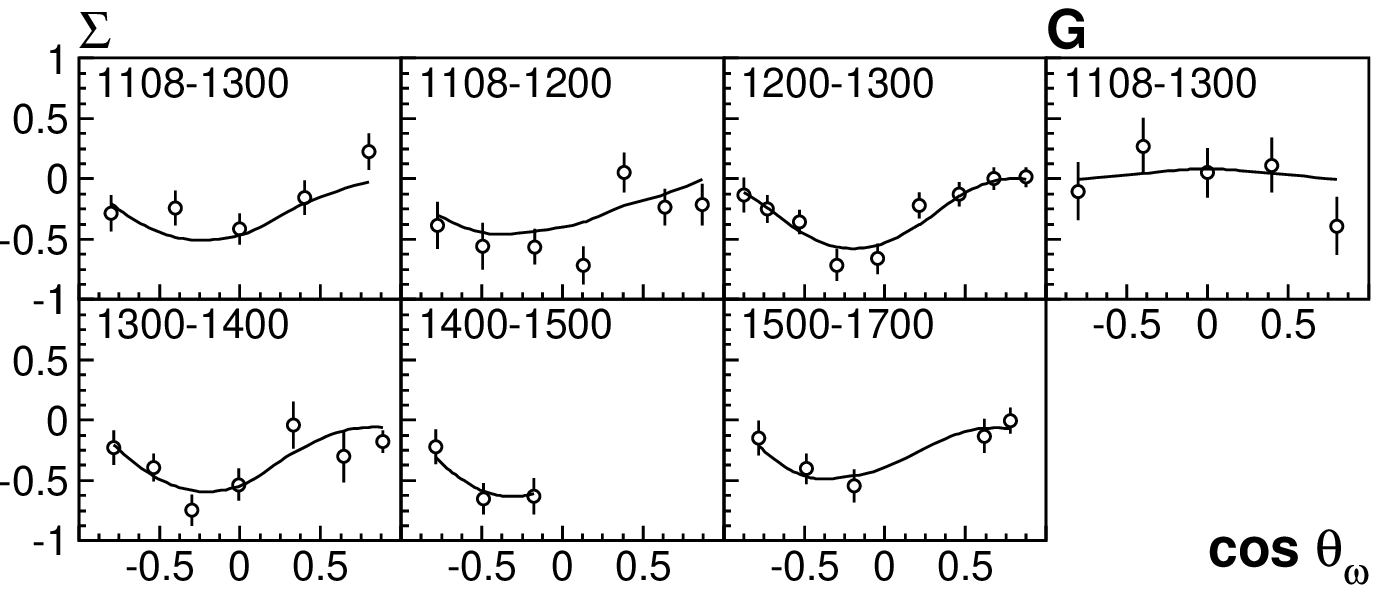}\\
\includegraphics[width=0.48\textwidth]{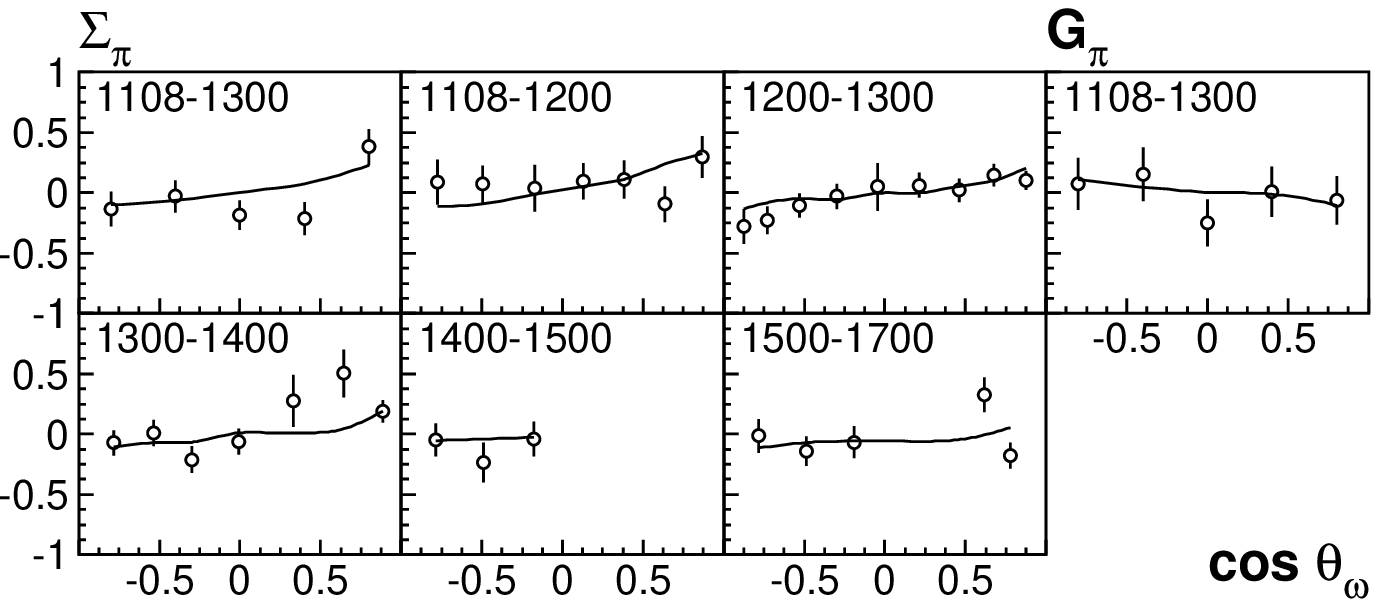}
\caption[dpol]{\label{fig:sigma}The
beam asymmetry $\Sigma$ (with respect
to $\omega$ direction) or $\Sigma_\pi$ (with respect to the direction of the $\pi^0$ from the
$\omega\to\pi^0\gamma$ decay) in bins of the photon energy \cite{Klein:2008aa}. 
The results on $G$, $G_\pi$ are from \cite{Eberhardt:2015lwa}.  }
\end{figure}

The differential cross sections, separated into 50~MeV
wide bins in incoming photon energy and 24 angular bins, are shown in
Figure~\ref{fig:Dcross_EF}. The distributions show a strong forward peaking, in particular 
at higher energies: diffractive production of $\omega$ mesons plays 
a role which becomes increasingly important with increasing photon energy. 
These and the other CBELSA/TAPS data are compared with the results of a partial wave analysis (PWA) fit described below.

\begin{figure}[pt]
 \centering
\includegraphics[width=0.48\textwidth]{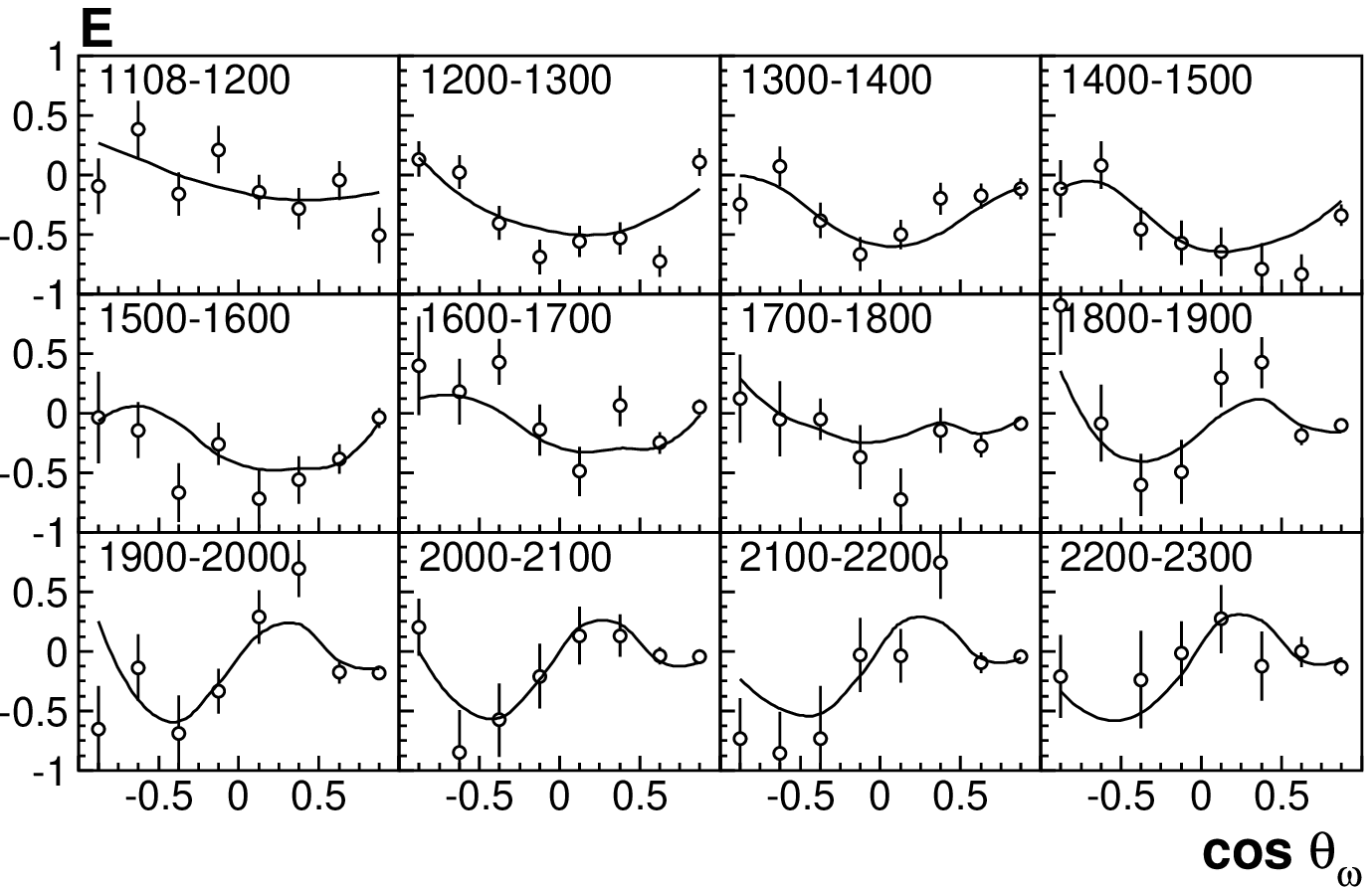}\\[-0.3ex]
\caption[dpol]{\label{fig:dpol}The
helicity asymmetry $E$ in bins of the photon energy \cite{Eberhardt:2015lwa}. \vspace{4mm} }
\centering
\includegraphics[width=0.48\textwidth,height=0.63\textheight]{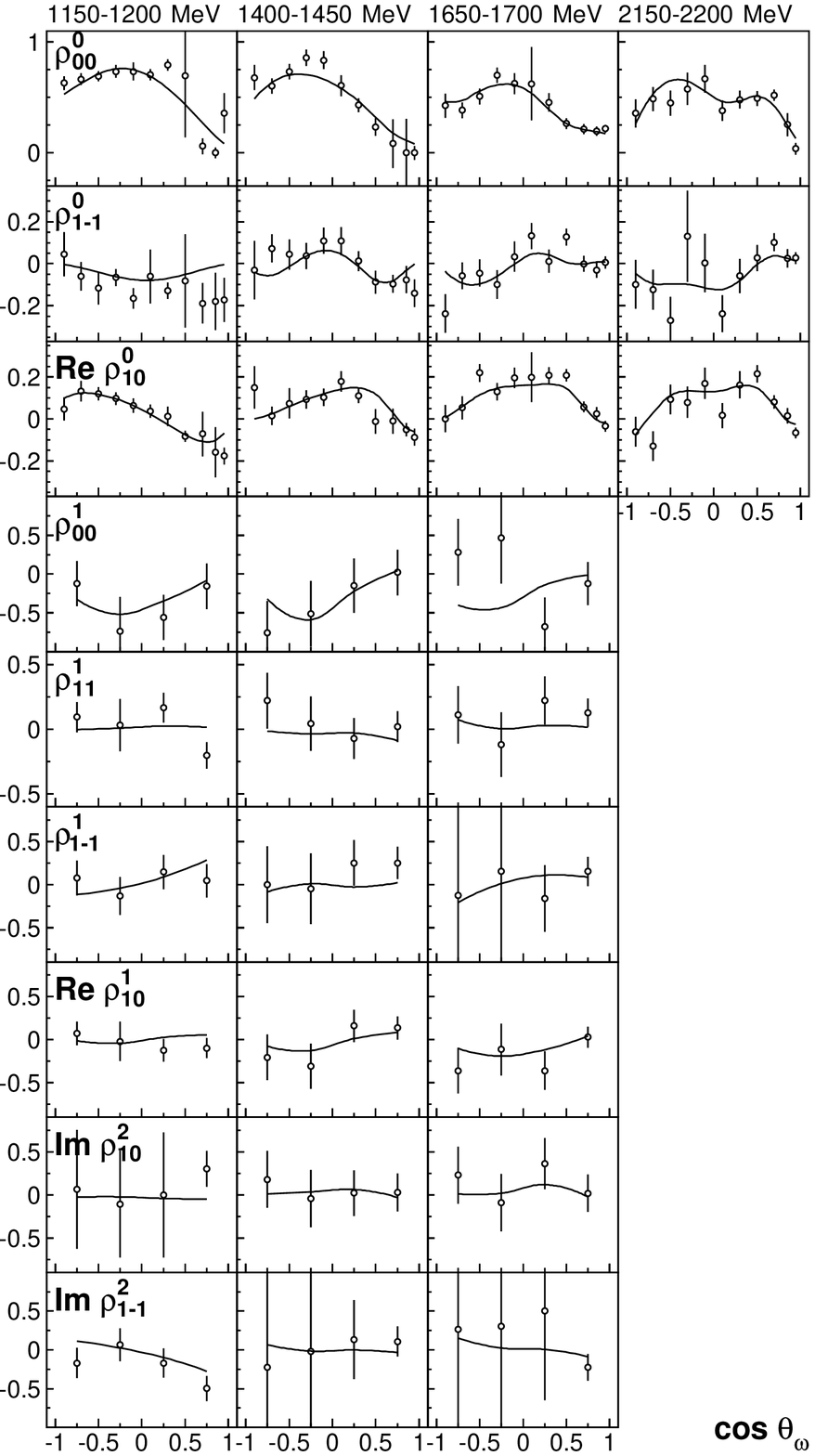}\vspace{-2mm}\\
\caption[SDME]{\label{fig:SDME}(Color Online)  Spin Density Matrix
Elements in the Adair frame from the CBELSA/TAPS
experiment for selected bins of the photon energy (in MeV) 
\cite{Wilson:2015uoa}. The total uncertainty
for each data point represented as a vertical bar. The solid curve
represents the BnGa PWA solution.  }
\end{figure}

The results on $\Sigma$ and $\Sigma_\pi$ are compared to the PWA
fit in Fig.~\ref{fig:sigma}. The results
have been reported earlier \cite{Klein:2008aa}. 
For the measurement of $E$ ($G$),  circularly (linearly) polarized
photons and longitudinally polarized protons were used. Data
selection and analysis are documented in \cite{Eberhardt:2015lwa}.
Here, the results on $G$ and $G_\pi$ are shown in Fig.~\ref{fig:sigma} and
those on $E$ in Fig.~\ref{fig:dpol}. The results are compared to the PWA fit.

A selection of
unpolarized SDMEs $\rho^0_{00}$, $\rho^0_{1-1}$, and
$Re\rho^0_{10}$ is shown in Figure~\ref{fig:SDME}. The events
were divided into 11 equally distributed angular bins. Also shown
in Figure~\ref{fig:SDME} are selected SDMEs
measured with linearly
polarized photons ($\rho^1$ and $\rho^2$): They
were extracted but now considering the polarization. Due
to the increased number of fit parameters (SDMEs) and the low statistics, the number of
angular bins was reduced to four equally
sized bins. The statistical uncertainties were estimated by
considering a large array of toy Monte Carlo generated data sets of
different sizes and polarization degrees. The systematic
uncertainties were found by considering experimental analysis
uncertainties, uncertainties from the Monte Carlo simulation, a
possible target shift away from the nominal position, and kinematic
fitting uncertainties \cite{Wilson:2015uoa}. The data not shown are 
fitted as well, with the same fit quality as the data shown in the figures.

\section{Partial wave analysis}
The data were included in the large BnGa data base covering pion and
photo-induced reactions. The fit uses the dispersion relation
approach based on the N/D technique which corresponds to the
solution of the Bethe-Sal\-peter equation in the case of a separable
interaction. In a simplified case when the regularization of the
dispersion integral is independent from the initial and final states
this method is algebraically equal to a modified K-matrix
approach:
 \be
 \mathbf{\hat A}(s) \;=\; \mathbf{\hat K}\;(\mathbf{\hat I}\;-
 \mathbf{\hat B \hat K})^{-1}\,. \label{k_matrix}
\ee
The multi-channel amplitude {\boldmath$\hat{{\rm A}}(s)$} with
the matrix elements ${A}_{ab}(s)$ defines the transition amplitude
from the K-matrix channel 'a' to the K-matrix channel 'b'.
$\mathbf{\hat B}$ is a diagonal matrix of the respective loop
diagrams with an imaginary part equal to the corresponding phase
space volume:
\be
\hat B_i= Re B_i+\;i\rho_i\,.
\ee
If the real part of the loop diagram is neglected, this method
corresponds to the classical K-matrix approach.

In the present fit we used a subtraction procedure to calculate
the elements of the B-matrix:
\bq
B_{i}(s)&=&b_i+(s-(m_{1i}+m_{2i})^2)\times\\
&&\hspace{-8mm}\int\limits_{(m_{1i}+m_{2i})^2}^\infty\!\!\!\! \frac{ds'}{\pi}
\frac{\rho_i(s')}{(s'-s-i\epsilon)(s'-(m_{1i}+m_{2i})^2)}\nonumber
\eq
where $\epsilon$ goes to zero. $b_i$ are subtraction constants,
and $m_{1i}, m_{2i}$ are
masses of the particles in channel $i$.

The K matrix elements combine the contributions from resonances
and from background:
\begin{equation}
\label{kmat}
K_{ab}=\sum\limits_\alpha \frac{g^{\alpha}_a
g^{\alpha}_b}{M_\alpha^2-s}+f_{ab}\,.
\end{equation}
Here $g^{\alpha}_{a,b}$ are coupling constants of the pole $\alpha$
to the initial state  $a$ and the final state $b$. The number of the
channels is varied for different partial waves. As a rule, it
includes two-body final states $\pi N, \eta N, K\Lambda$
\cite{Anisovich:2011fc}, $K\Sigma$ \cite{Anisovich:2013tij} and a
number of intermediate mesonic and baryonic resonances which
contribute notably to the $\gamma p\to\pi^0\pi^0 p$ \cite{Sokhoyan:2015fra} and
$\gamma p\to\pi^0\eta p$ \cite{Gutz:2014wit} cross sections. In addition, we
include into the K-matrix one additional channel which
describes the contribution from channels which are not taken
into account explicitly. In the present solution the phase volume of
inelastic channel was parametrized as the $\rho N$ ($\omega N$)
contribution with the lowest possible orbital angular momentum.

The amplitude (\ref{k_matrix}) corresponds to the sum of the tree
level diagrams described by the K-matrix and diagrams with
consequent rescattering due to loop diagrams (defined by the
K-matrix channels) and vertices defined by the K-matrix. In the case
of the photoproduction amplitude the initial $\gamma N$ interaction
is taken into account only once and neglected in the
rescattering loops due to a small coupling constant. This approach
is called the P-vector approach (see e.g \cite{Chung:1995dx}):
\bq\label{eq5a}
a^h_{b}&=P^h_a(I-BK)_{ab}^{-1}\qquad {\rm where}\\
&P^h_a=\sum\limits_\alpha \frac{A^h_{\alpha}\nonumber
g^{\alpha}_a}{M_\alpha^2-s}+F_{a}.
\eq
Here $A^h_{\alpha}$ is the helicity-dependent photo-coupling of a
pole $\alpha$ and $F_{a}$ a non-resonant transition.

It is also useful to rewrite the sum of rescattering diagrams
extracting the transition to the final state:
\bq\label{Dvect}
A_{af} &=& \hat D_{af} + [\hat K (\hat I\!-\!\hat B \hat
K)^{-1}\,\hat B ]_{ab} \hat D_{bf}\\
D_{bf} &=& \sum_\alpha \frac{g_b^{(\alpha)}g_{f}^{(\alpha)} }{M^2_\alpha - s} \;+\;
 \tilde d_{bf}\,.~~
 \eq
Here $g_{f}^{(\alpha)}$ is the coupling of a resonance to the final
state and $\tilde d_{bf}$ represents the non-resonant transition from
the K-matrix channel $b$ to the final state $f$. If the final state
corresponds to the one of the K-matrix channels, the
amplitude (\ref{Dvect}) will be the same as the amplitude (\ref{k_matrix}).
However, this expression allows us to describe the transition to weak
channels.

In cases where both, initial and final coupling constants, are weak
we use an approximation which we call PD-vector. In this case the
amplitude is given by
\be
A_{f} \;=\; \hat G_{f} + \hat P_{a}[(\hat I\;-\;\hat B \hat
K)^{-1}\,\hat B ]_{ab} \hat D_{bf}\;\;.
\ee
$\hat G_{f}$ corresponds to a tree diagram for the transition from
initial channel ($\gamma N$ in the case photoproduction)  to the
state '$f$':
\be
 G_{f}\;=\;\sum_\alpha \frac{g_{\gamma N}^{(\alpha)}g_{f}^{(\alpha)} }{M^2_\alpha - s} \;+\;
 \tilde h_{(\gamma N)f}\,.
 \label{PDvect}
 \ee
Here, the elements $\tilde h_{(\gamma N)f}$ represent the direct non-resonant transitions
from the initial photon-nucleon system to the different final
states. These are the only new parameters of the fit once the P-vector
and D-matrix are known. In the present analysis we did not introduce
this non-resonant transition; instead we included reggeized pion
and Pomeron exchange amplitudes. These are represented by the
exchange of a Reggeon \cite{Sarantsev:2008ar} in the form
\bq
A&=&g(t)R(\xi,\nu,t) \quad {\rm where}\\
R(\xi,\nu,t)&=&\frac{1+\xi
exp(-i\pi\alpha(t))}{\sin(\pi\alpha(t))} \left (\frac{\nu}{\nu_0}
\right )^{\alpha(t)} \;.\nonumber
\eq
We use $g(t)=g_0\exp(-bt)$  as vertex function and form factor.
$\alpha(t)$  describes the trajectory, $\nu=\frac 12 (s-u)$, $\nu_0$
is a normalization factor, and $\xi$ the signature of the
trajectory. Pion and and Pomeron exchange both have a positive
signature and therefore \cite{Anisovich:2004zz}:
\be\label{eq9a}
R(+,\nu,t)=\frac{e^{-i\frac{\pi}{2}\alpha(t)}} {\sin
(\frac{\pi}{2}\alpha(t))} \left (\frac{\nu}{\nu_0}\right
)^{\alpha(t)}\;.
\ee

To eliminate the poles at $t<0$, additional $\Gamma$-functions are
introduced in (\ref{eq9a}).
\be
\sin \left (\frac{\pi}{2}\alpha(t)\right ) \to \sin \left
(\frac{\pi}{2}\alpha(t)\right )  \; \Gamma \left (\frac
{\alpha(t)}{2}\right )\, .
\label{rho_1}
\ee
The pion and Pomeron trajectories were taken with the standard
parameterization:
\bq\label{eq11a}
\hspace{-5mm}\pi\quad\qquad\qquad\alpha(t)&=-&0.25+0.85 ({\rm GeV}^{-2})t\\
\label{eq11c}
\hspace{-5mm}Pomeron\quad\quad\alpha(t)&=&0.26+0.85  ({\rm GeV}^{-2})t
\eq
where $t$ should be given in GeV$^2$. 

The amplitude in the form of eq.~(\ref{PDvect}) is very suitable for the
description of reactions with a relatively small cross section.
In this case the new parameters describing the decay of
resonances into the new channel or the non-resonant transitions do
not influence the description of other reactions. For example, in
the case of one resonance, the amplitude (\ref{PDvect}) corresponds
to a relativistic Breit-Wigner amplitude with production and decay
couplings in the numerator and a resonance width formed by the
K-matrix channels.

\section{Fit results}
The fit with only the t-channel exchange amplitudes demonstrates
clearly the importance of the matrix-density data and of the polarization
observables. For example, a fit with only Pomeron exchange
reproduces well the differential cross section above 2000\,MeV, but
predicts vanishing $\rho_{00}$ density matrix elements, a vanishing beam asymmetry,
a vanishing helicity asymmetry,  and a vanishing $G$-observable. A fit which includes
Pomeron and pion exchanges predicts $\rho_{00}$ and the beam asymmetry to be very close to zero.

In our first fits, the $N\omega$ decays were admitted for all known
$N^*$ resonances \cite{Agashe:2014kda} above or just below
the $N\omega$ threshold. The reaction was fitted using the
PD-vector approach (\ref{PDvect}) which allows us to use directly
solution BG2014-02 with fixed parameters. We tried about 200
different fits starting from different initial couplings. The best
solution showed a large contribution from the $J^P=3/2^+$ partial
wave already just above the reaction threshold. The partial waves
$J^P=3/2^-$ and $J^P=1/2^-$ which can couple to the $N\omega$
channel in the $S$-wave are considerably smaller. In the energy range
considered here, the Pomeron-exchange contribution rises
continuously with energy and reaches about 50\% of the total cross
section at $W=2000$\,MeV. The pion exchange contribution was found to
be small, although the fit quality hardly changes if one enforces it to be up
to 20\% of the total cross section.

The best fit with the known $N^*$ states provides a good description
of the data for masses below 2100\,MeV, and an acceptable description
above. We tried to improve the description by adding Breit-Wigner
resonances with different quantum numbers. The best improvement was
obtained when an additional resonance was introduced with a mass
above 2200\,MeV. Its mass optimized around 2230\,MeV; its quantum
numbers are not well defined:
 $J^P= 1/2^-,3/2^+$, $3/2^-$, or $5/2^+$ lead to a similar fit quality.
Although the new state influences the parameters of lower mass resonances only
slightly, it provides some flexibility of the fit in the lower mass region and
led to a significant improvement in the description of the density
matrix elements also below 2100\,MeV.

Next, we investigated the stability of the solution by excluding one by one the $\omega
N$ couplings of the resonances. Some couplings could be put
to zero, with an almost negligible deterioration of the fit. These
solutions were included in the systematic error estimation.

\begin{table}[pt]
\caption{\label{chi} $\chi^2$ for the solution with the standard set
of $N^*$ resonances and for the solution where one resonance with $J^P=1/2^-$
at about 2230\,MeV is added.  }
\renewcommand{\arraystretch}{1.3}
\bc\footnotesize
\begin{tabular}{ccccccccc}
\hline\hline 
\hspace{-1mm}$N(2230)$\hspace{-3mm}&\hspace{-3mm} $d\sigma/d\Omega$ \hspace{-3mm}&\hspace{-3mm} $\Sigma$ \hspace{-3mm}&\hspace{-3mm} $\Sigma_\pi$ \hspace{-3mm}&\hspace{-3mm}
$G$+$G_\pi$ \hspace{-3mm}&\hspace{-3mm} E \hspace{-3mm}&\hspace{-3mm} $\rho_{00}$ \hspace{-3mm}&\hspace{-3mm} $\rho_{10}$ \hspace{-3mm}&\hspace{-3mm} $\rho_{1-1}$ \\[0.5ex]
\hline No  \hspace{-3mm}&\hspace{-3mm} 0.72 \hspace{-3mm}&\hspace{-3mm} 0.85 \hspace{-3mm}&\hspace{-3mm} 1.20 \hspace{-3mm}&\hspace{-3mm} 0.59  \hspace{-3mm}&\hspace{-3mm}
1.29 \hspace{-3mm}&\hspace{-3mm} 1.42 \hspace{-3mm}&\hspace{-3mm}1.23 \hspace{-3mm}&\hspace{-3mm} 1.36\\
Yes \hspace{-3mm}&\hspace{-3mm} 0.56 \hspace{-3mm}&\hspace{-3mm} 0.85 \hspace{-3mm}&\hspace{-3mm} 1.19 \hspace{-3mm}&\hspace{-3mm} 0.58  \hspace{-3mm}&\hspace{-3mm} 1.16 \hspace{-3mm}&\hspace{-3mm} 1.10 \hspace{-3mm}&\hspace{-3mm}1.04 \hspace{-3mm}&\hspace{-3mm} 1.29 \\
\hline
N$_{\rm data}$   \hspace{-2mm}&\hspace{-2mm} 648 \hspace{-2mm}&\hspace{-2mm} 36 \hspace{-2mm}&\hspace{-2mm} 36 \hspace{-2mm}&\hspace{-2mm} 10  \hspace{-2mm}&\hspace{-2mm} 95 \hspace{-2mm}&\hspace{-2mm} 297 \hspace{-2mm}&\hspace{-2mm}297 \hspace{-2mm}&\hspace{-2mm} 297 \\
\hline\hline
\end{tabular}\vspace{-3mm}
\ec
\end{table}

At the next step we included in the K-matrix those $N\omega$ channels 
which provided significant contributions to the partial wave (instead of
treating them as PD vectors). In many partial waves,
the fit just reduced the partial width of the {\it missing channel}
in favor of the $N\omega$ channel. In some cases, however,  we
had to refit the whole data base to find an improved solution.

The final refit of the Bonn-Gatchina data base, with the data on
$\gamma p\to \omega p$ included, produced almost the same quality of
the description of the other data sets: neither pole positions or
decay properties of the resonances changed significantly. However, a
small tuning of all couplings allowed us to improve notably the
description of the $\gamma p\to \omega p$ observables. 

The K-matrix solution where a $J^P=1/2^-$ resonance in the region 
2230 MeV was admitted in the fit was taken as main solution.
All other solutions were used to estimate the
errors from the range of values obtained in the other fits. 
In Table~\ref{chi} we give the breakdown of the $\chi^2$
contributions.  As an example, we show the $\chi^2$ values when the high
mass $J^P=1/2^-$ resonance is excluded from the fit. The fit quality is
very similar when the spin-parity is changed to $3/2^+$, $3/2^-$, or $5/2^+$. 

Table~\ref{BR} lists those resonances which have an $N\omega$ decay mode
which yields a significant
improvement of the fit quality. The table gives the branching
ratios, their errors, and the change in $\chi^2$ when the coupling
of a resonance is fixed to zero. The results reported in \cite{Penner:2002ma}
are listed as small numbers.

A few comments need to be made:
\begin{table}[pt]
\caption{\label{BR} Branching ratios (B.R. in \%) for $N^*$ decays
into $N\omega$. Small numbers were reported in \cite{Penner:2002ma}.
The $\delta(\chi^2)$ values give the change in $\chi^2$ when the
$N\omega$ decay mode is excluded.\vspace{-2mm}}
\renewcommand{\arraystretch}{1.3}
\bc\footnotesize
\begin{tabular}{lcclcc}
\hline\hline Resonance & B.R. & $\delta(\chi^2)$ &
Resonance & B.R. & $\delta(\chi^2)$ \\
\hline
$N(1700){3/2^-}$   & 22\er12 &100 &$N(1900){3/2^+}$&15\er 8&  70 \\[-1.6ex]
&&&&\tiny  13\er9&\\[-0.5ex]
$N(1710){1/2^+}$   &  2\er2  & 26 &$N(2000){5/2^+}$&18\er 8&  42 \\[-1.6ex]
&\tiny  8\er5&&&\tiny  1\er1&\\[-0.5ex]
$N(1720){3/2^+}$   & 26\er 14&105 &$N(2060){5/2^-}$&4\er3  &  37 \\
$N(1875){3/2^-}$   & 13\er 7 & 98 &$N(2100){1/2^+}$&15\er10&  78 \\[-1.6ex]
&\tiny 20\er4&&&\\[-0.5ex]
$N(1880){1/2^+}$   & 20\er 8 & 33 &$N(2150){3/2^-}$&12\er 8&  99 \\
$N(1895){1/2^-}$   & 28\er12 &100 &$N(2190){7/2^-}$&14\er 6& 131 \\
\hline\hline
\end{tabular}\vspace{-3mm}
\ec
\end{table}

From the data on $\gamma p\to \omega p$ alone, only the products
of the helicity amplitudes $A_{1/2}$, $A_{3/2}$ and the square root
of the $N^*\to N\omega$ branching ratios are determined. The helicity
amplitudes can be deduced when elastic scattering data and photoproduction
of pions are included in the fits.   

The comparison of our results with those of Penner and Mosel
\cite{Penner:2002ma} shows good consistency. Only the $N\omega$ branching
ratios of the $N(2000)5/2^+$ resonance are different in magnitude. 
However, the $N(2000)5/2^+$ resonance is difficult to observe in photoproduction,
and our result is only slightly more than 2$\sigma$ away from zero.  

The pole positions  of $N(1700){3/2^-}$  is fitted to values just
above the $N\omega$ threshold (at 1720\,MeV), those of
$N(1680){5/2^+}$, $N(1710){1/2^+}$ and $N(1720){3/2^+}$ below the
threshold. The $N\omega$ coupling constants of these states are
non-zero, leading to a non-vanishing amplitude above the $N\omega$
threshold, and suppressing these couplings leads to a notable
deterioration of the fit quality. Formally, the branching ratios
would be vanishing or very small since the phase space at the
nominal mass of the resonance is zero or small. Therefore, the
branching ratios of Table~\ref{BR} for these resonances were
calculated by numerical integration over the full width of the
resonance. For higher-mass resonances, the numerical integration or
the usual definition of the branching ratio give nearly identical
results.  The $N(1900){3/2^+}$ in Table~\ref{BR} resonance stands
for a complex of two resonances - suggested as well in
\cite{Williams:2009ab,Williams:2009aa} - but only the properties of
$N(1900){3/2^+}$ are well defined in our fits.

The branching ratios are derived from the best ten fits which yield
acceptable descriptions of the data. The spread of their results is used to define the errors.
The spread is hence due to systematic uncertainties, the statistical
uncertainties are small. Note that the errors can be large even in
cases where the statistical significance is high.

\section{Summary}
In summary, we have reported a partial wave analysis including
new data on the reaction $\gamma p\to \omega p$
for unpolarized and polarized photons and unpolarized and polarized
protons. The analysis is performed within the Bonn-Gatchina partial
wave formalism and includes other data on pion and photo-induced
reactions. Branching ratios of twelve nucleon  resonances
for their decay into nucleon plus $\omega$ are derived.

{\small We would like to thank the members of the CBELSA/TAPS collaboration
for letting us use their data before publication. 
We acknowledge support from the \textit{Deutsche
Forschungsgemeinschaft} (SFB/TR16), \textit{Russian Foundation
for Basic Research}, and \textit{ U.S. National Science Foundation}.}

\end{document}